\documentclass[12pt, a4paper] {article}

\usepackage{amsfonts}
\usepackage{amsmath}
\usepackage{verbatim}
\usepackage{color} 
\usepackage{epsfig}        
 \usepackage{graphicx}
\begin{document}

\begin{center}
 
{\bf {\large Novel Symmetries in Vector Schwinger Model}}\\
\vskip 1 cm

{\bf Saurabh Gupta }\\
{{\it The Institute of Mathematical Sciences, \\ CIT Campus,  Chennai - 600 113, India}}\\
{\small {\bf e-mail: saurabh@imsc.res.in}}

\end{center}

\vskip 1 cm

\noindent
{\bf Abstract:} We derive nilpotent and absolutely anticommuting (anti-)co-BRST symmetry transformations for the 
bosonized version of  $(1+1)$- dimensional (2D) vector Schwinger model. These symmetry transformations turn out 
to be the analogue of co-exterior derivative of differential geometry as the total gauge-fixing term remains 
invariant under it. The exterior derivative is realized in terms of the (anti-)BRST symmetry transformations of the theory
whereas the bosonic symmetries find their analogue in the Laplacian operator. 
The algebra obeyed by these symmetry transformations turns out to be exactly same as the algebra 
obeyed by the de Rham cohomological operators of differential geometry. 
 
\vskip 1.5 cm



\newpage

\noindent 
{\bf 1. Introduction}\\

\noindent 
The Becchi-Rouet-Stora-Tyutin (BRST) formalism is one of the most intuitive approaches to quantize a gauge theory. In BRST
formalism the unitarity and ``quantum" gauge (i.e. BRST) invariance are respected together at any arbitrary order of 
perturbative computations. These (anti-)BRST symmetry transformations always satisfy the two sacrosanct properties: 
(i) the nilpotency of order two, and (ii) the absolute anticommutativity. The former property implies fermionic nature of 
the (anti-)BRST symmetry transformations whereas the latter one encodes the linear independence of these transformations 
(see, e.g. \cite{brs1,brs2,brs3,tyu}).

The Schwinger model, which describes the quantum electrodynamics in $(1+1)-$dimension with massless fermions, is a well-studied 
model as far as the two dimensional field theories are concerned (see, e.g. \cite{sch1,suss,suss1,halp,boyan,raja1,raja2,flack,malik}). 
In this work we consider the $(1+1)-$dimensional bosonized version of vector Schwinger model (VSM). The VSM (as well as the chiral 
Schwinger model) can be obtained from a generalized version of Schwinger model (see, e.g. \cite{boyan} for details). The VSM 
is an exactly solvable model and endowed with the first-class constraints, in the language of Dirac's prescription for the classification 
of constrained systems, which makes it a gauge invariant model. The Hamiltonian and BRST formulations of this model have been
discussed in \cite{usha}.

As far as the framework of BRST formalism is concerned, it has been shown that any $p-$form (with $p = 1,2,3$) Abelian gauge theories in $D = 2p$ 
dimensions of spacetime are tractable models of Hodge theory \cite{sght,sgcsm,malik1,rkma3,mal2p}, where 
the underlying theory is endowed with, in totally, six continuous symmetries [i.e. (anti-)BRST, (anti-)co-BRST, bosonic and ghost symmetries].
At this juncture, it is worthwhile to mention that the higher $p-$form $(p \geq 2 )$ fields appear in the excitations of the quantized versions 
of (super)strings and related extended objects (see, e.g. \cite{green}). 
Moreover, at algebraic level, the 1D model of rigid rotor also provides a toy model for Hodge theory \cite{sgrr}. 
With the help of such kind of studies it has been proven  that 2D Abelian 1-form gauge theory is a new model for topological field theory \cite{mal01}, 
whereas 4D Abelian 2-form gauge theory provides an example of quasi-topological field theory \cite{mal03}. 
Thus, these kind of studies play an important role from physical point of view. 
Furthermore, in the case of non-Abelian gauge theories, it has been shown that the 2D free non-Abelian 1-form gauge theory provides a model for Hodge theory \cite{mal04}.

The prime motivation towards the present investigation comes from one of our recent works \cite{sgcsm} on the chiral Schwinger model (CSM)
where we have shown that the modified version of 2D bosonized CSM is endowed with, in totality, six [i.e. (anti-)BRST, (anti-)co-BRST, 
bosonic and ghost] continuous symmetry transformations. Furthermore, this model has been shown to be a tractable field theoretic model 
for the Hodge theory where all the de Rham cohomological operators of differential geometry find their analogue in terms of the symmetry 
transformations (and their corresponding generators) of the underlying theory. Thus, keeping above in mind, it is worthwhile to investigate whether  
the bosonized version of 2D VSM have the similar kind of symmetry structure as that of the modified version of 2D CSM.  We find, within
the framework of BRST formalism, that both the above mentioned models have similar properties as far as the continuous symmetries and 
their algebraic structures are concerned.

To establish the existence of (anti-)co-BRST symmetries, in this model, is also important due to the following reasons. First, the (anti-)BRST and (anti-)co-BRST 
symmetries have completely different origins and realized in different ways (see, for details \cite{malik1}). The different way of 
realization implies that the co-BRST symmetries can give different superselection sector from the BRST symmetries \cite{yang}. Second, the physical states of the 
underlying theory could be locally identified with those states that are {\it both} (anti-)BRST and (anti-)co-BRST invariant. Thus, for the direct cohomological 
description of the physical states of the system, only BRST charge (corresponding to the BRST symmetry) is not enough \cite{mcmu}.

Our present paper is organized as follows. In the second section, we briefly discuss about the gauge symmetries, constrained 
structure and first-order formalism of the 2D bosonized version of VSM for the sake of completeness. The third section contains 
a discussion about the off-shell
nilpotent and absolutely anticommuting (anti-)BRST symmetry transformations. Our fourth section is devoted to the derivation of 
the off-shell nilpotent (anti-)co-BRST symmetry transformations. In the fifth section, we derive a bosonic symmetry transformations. 
Our sixth section includes the discussion on ghost and discrete symmetries of the theory. 
The algebraic structure obeyed by the above symmetry transformations and their connection with the de Rham cohomological operators 
of differential geometry is included in seventh section. Finally, in the last section, we make concluding remarks and point out 
some future directions. \\

\noindent 
{\bf 2. Preliminaries: Gauge symmetries }\\

\noindent 
We start with the following Lagrangian density of $(1+1)-$dimensional bosonized version of vector Schwinger model \cite{boyan}
\begin{eqnarray}
 {\cal L}_{VSM} &=& - \frac{1}{4} \; F_{\mu\nu} F^{\mu\nu} - e \; \varepsilon^{\mu\nu} \partial_\mu \phi \; A_\nu
+ \frac{1}{2} \; \partial_\mu \phi \; \partial^\mu \phi \nonumber\\ \label{1}
&\equiv& \frac{1}{2} E^2 - e \; \varepsilon^{\mu\nu} \partial_\mu \phi \; A_\nu
+ \frac{1}{2} \; \partial_\mu \phi \; \partial^\mu \phi, 
\end{eqnarray}
where $F_{\mu\nu} = \partial_\mu A_\nu - \partial_\nu A_\mu$ is field strength tensor for 1-form $(A^{(1)} = dx^\mu A_\mu )$ gauge field $A_\mu$.
In the above\footnote{We choose here the 2D flat metric $\eta_{\mu\nu}$ 
with signature $(+1,-1)$ where 
the Greek indices $\mu, \nu ... = 0, 1$. The 2D Levi-Civita tensor $\epsilon_{\mu\nu} = - \epsilon_{\nu\mu}$ is such that 
$\varepsilon_{01} = +1 = - \varepsilon^{01}$ and it obeys $\epsilon^{\mu\nu} \epsilon_{\mu\nu} = - 2 !, \; 
\epsilon^{\mu\nu} \epsilon_{\mu\kappa} = - \delta^\nu_\kappa$, etc. The overdot and prime indicate the time and space derivatives, respectively
(i.e. $\dot A = \frac {dA}{dt}$ and $B^\prime = \frac{dB}{dx}$).}, the first term represents the kinetic energy term for the gauge field $A_\mu$ and, 
in 2D, it has only electric field $E$ as its existing component. Moreover, there are no propagating degrees of freedom left for $A_\mu$ in 2D. 
The second term corresponds to the coupling of gauge field with massless bosonic field $\phi$ (or equivalently fermionic field in 2D) where $e$ is a 
coupling constant. The last term is the kinetic term for field $\phi$. The canonical conjugate momenta, calculated from above Lagrangian 
density, are:
\begin{eqnarray}
 \Pi_\phi = \dot \phi + e A_1, \qquad \Pi^0 = 0,  \qquad \Pi^1 = E. \label{2}
\end{eqnarray}
It is clear from the above that $(\chi_1 := )\Pi^0 \approx 0 $ is a primary constraint on the theory and by demanding that the primary constraint 
should remain intact with respect to time leads to the secondary constraint  ($\chi_2 :=) E^\prime - e \phi^\prime \approx 0$. It is straightforward
to check that there are no further constraints in the theory \cite{usha}. It is turn out that, using Dirac's prescription for the classification of 
constraints \cite{dirac,sunder}, that the above mentioned constraints  $\chi_1$ and $\chi_2 $ are first-class in nature. This implies that the underlying theory
is a gauge theory. The canonical Hamiltonian density (${\cal H}_c$), calculated from (\ref{1}) and (\ref{2}), has following structure:
\begin{eqnarray}
 {\cal H}_c &=& \Pi_\phi \dot \phi + \Pi_0 \dot A_0 + E \dot A_1 - {\cal L}_{VSM} \nonumber\\
&=& \frac{1}{2} (E^2 + \Pi_\phi^2 + \phi^{\prime 2} + e^2 A_1^2) + E A_0^\prime
- e \Pi_\phi A_1 + e \phi^\prime A_0.
\end{eqnarray}
Therefore, the total Hamiltonian density (${\cal H}_T$) can be given as
\begin{eqnarray}
 {\cal H}_T = {\cal H}_c + \Pi_0 \; \lambda, 
\end{eqnarray}
where $\lambda$ is a Lagrange multiplier and $\Pi_0$ is the primary constraint on the theory. Thus, the first-order Lagrangian density\footnote{We  
differ from the first-order Lagrangian density of \cite{usha} for the sake of brevity and algebraic convenience.} (${\cal L}_F$), has following form \cite{usha}
\begin{eqnarray}
 {\cal L}_F &=& \frac{1}{2} \Big(E^2 -\Pi_\phi^2 - \phi^{\prime 2} - e^2 A_1^2 \Big) + \Pi_\phi \; \dot \phi 
+ e \; \Pi_\phi A_1 - e \; \phi^\prime A_0  + p_\lambda \dot \lambda,  \label{3}
\end{eqnarray}
here $p_\lambda$ is canonically conjugate momenta to $\lambda$. The above mentioned first-order Lagrangian density remains invariant under following 
infinitesimal gauge symmetry transformation $(\delta_g)$ 
\begin{eqnarray}
&& \delta_g A_0  = \dot \beta, \quad \delta_g A_1 = \beta^\prime, \quad \delta_g \Pi_\phi = e \beta^\prime, \quad \delta_g E = 0, \nonumber\\
&& \delta_g \phi = 0,  \qquad \delta_g \lambda = 0, \qquad \delta_g p_\lambda = 0,  \label{6}
\end{eqnarray}
because ${\cal L}_F$ goes to a total spacetime derivative, as 
\begin{eqnarray}
 \delta_g {\cal L}_F =   \partial_\mu [ e \beta \; \varepsilon ^{\mu\nu}  \; \partial_\nu \phi].
\end{eqnarray}
Therefore, the corresponding action remains invariant and hence (\ref{6}) are the symmetry transformations of the theory. \\

\noindent
{\bf 3. (Anti-)BRST symmetry transformations: Analogue of exterior derivative}\\

\noindent
The (anti-)BRST invariant first-order Lagrangian 
density, in its full blaze of glory, can be given as follows: 
\begin{eqnarray}
 {\cal L}_b &=&  \frac{1}{2} (E^2 - \Pi_\phi^2 - \phi^{\prime 2} - e^2 A_1^2) + \Pi_\phi \dot \phi +  e \Pi_\phi A_1 
- e \phi^\prime A_0  + p_\lambda \dot \lambda \nonumber\\
&+& b(\dot A_0 - A_1^\prime) + \frac{b^2}{2} + \dot{\bar C} \dot C - \bar C^\prime C^\prime, \label{lagb}
\end{eqnarray}
where $(\bar C) C$ are fermionic $[C^2 = \bar C^2 = 0,\; C \bar C + \bar C C = 0]$(anti-)ghost fields and $b$ is the Nakanishi-Lautrup 
auxiliary field which is used to linearize the gauge fixing term.  The following off-shell nilpotent (anti-)BRST symmetry transformations ($s_{(a)b}$)
\begin{eqnarray}
&& s_b A_0 = \dot C, \quad s_b A_1 = C^\prime, \quad s_b C = 0, \quad s_b {\bar C} = b, \nonumber\\
&& s_b \Pi_\phi = e C^\prime, \quad s_b E = 0, \quad s_b [b, \; \phi, \; p_\lambda, \; \lambda]  = 0, \nonumber\\
&& s_{ab} A_0 = \dot {\bar C}, \quad s_{ab} A_1 = \bar C^\prime, \quad s_{ab} \bar C = 0, \quad s_{ab} C = - b, \nonumber\\
&& s_{ab} \Pi_\phi = e {\bar C}^\prime, \quad s_{ab} E = 0, \quad s_{ab} [b, \; \phi, \; p_\lambda, \; \lambda]  = 0, \label{7}
\end{eqnarray}
leave the Lagrangian density (\ref{lagb}) quasi-invariant as obvious from the expressions given below 
\begin{eqnarray}
s_b {\cal L}_b = \partial_\mu [e \; C \; \varepsilon^{\mu\nu} \; \partial_\nu \phi + b \;\partial^\mu C], \quad
s_{ab} {\cal L}_b = \partial_\mu [e \; {\bar C} \; \varepsilon^{\mu\nu} \; \partial_\nu \phi + b \;\partial^\mu {\bar C}]. 
\end{eqnarray}
It is worthwhile to mention that the kinetic term (i.e. $\frac{1}{2} E^2$) remains invariant under (anti-)BRST symmetry transformations as 
it is evident from (\ref{7}) that $s_{(a)b}[E] = 0$. This kinetic 
term (i.e. $- \frac{1}{4} F_{\mu\nu} F^{\mu\nu} = \frac{1}{2} E^2$) owes its origin to the exterior derivative $d = dx^\mu \partial_\mu$
(with $d^2 = 0$) because the two-form $B^{(2)} = \frac{1}{2!} (dx^\mu \wedge dx^\nu) F_{\mu\nu}$ defines it through $B^{(2)} = d A^{(1)}$
where $A^{(1)} = dx^\mu A_\mu$ introduces the gauge potential $A_\mu$.

At this juncture, it is interesting to point out that the gauge fixing and ghost terms of above (anti-)BRST invariant
Lagrangian density can be derived in the following standard fashion (modulo a total derivative), using \cite{nem}
\begin{eqnarray}
s_b \; \Big[ \bar C \Big\{ (\partial \cdot A) + \frac {b}{2} \Big \} \Big] &=& b (\dot A_0 - A_1^\prime) + \frac{b^2}{2} 
+ \dot{\bar C} \dot C - \bar C^\prime C^\prime. \label{9}
\end{eqnarray}
The above gauge fixing condition (\ref{9}) can be, equivalently (modulo a total derivative), written as 
\begin{eqnarray}
s_{ab}  \Big[- C \Big\{ (\partial \cdot A) + \frac {b}{2} \Big \} \Big] &=&  b (\dot A_0 - A_1^\prime) + \frac{b^2}{2} 
+ \dot{\bar C} \dot C - \bar C^\prime C^\prime \nonumber\\
& \equiv & s_{ab} s_b \Big[ \frac{A^\mu A_\mu}{2} + \frac{C \bar C}{2}\Big].\label{10}
\end{eqnarray}
The relationship between (\ref{9}) and (\ref{10}) holds good because of the absolutely anticommuting nature (i.e. $s_b s_{ab} + s_{ab} s_b = 0 $) 
 of the (anti-)BRST symmetry transformations (\ref{7}). It is straightforward to check that 
$(s_b s_{ab} + s_{ab} s_b) \Phi = 0$,  where $\Phi (= A_\mu, C, \bar C, \phi, \lambda, b)$ is the generic field of (\ref{lagb}).\\

\noindent
{\bf 4. (Anti-)co-BRST symmetry transformations: Analogue of co-exterior derivative} \\

\noindent
The (anti-)BRST invariant Lagrangian density ${\cal L}_b$ is also endowed with the off-shell nilpotent $[s_{(a)d}^2 = 0]$
(anti-)co-BRST symmetry transformations $s_{(a)d}$. For this purpose, we incorporate an auxiliary field $\bar b$ to linearize 
the kinetic term as below: 
\begin{eqnarray}
 {\cal L}_d &=&  {\bar b} E - \frac{1}{2} ({\bar b}^2 + \Pi_\phi^2 + \phi^{\prime 2} +  e^2 A_1^2 ) + \Pi_\phi \dot \phi +  e \Pi_\phi A_1
-e \phi^\prime A_0 \nonumber\\ 
&+& p_\lambda \dot \lambda + b(\dot A_0 - A_1^\prime) + \frac{b^2}{2} + \dot{\bar C} \dot C - \bar C^\prime C^\prime. \label{11}
\end{eqnarray}
The above Lagrangian density respects following off-shell nilpotent (anti-)co-BRST symmetry transformations
\begin{eqnarray}
&& s_d A_0 = - {\bar C}^\prime, \qquad s_d A_1 = - \dot {\bar C}, \qquad s_d \Pi_\phi = - e \dot {\bar C}, \qquad s_d E = - \Box \bar C, \nonumber\\
&& s_d C = \bar b - e \phi, \qquad s_d \bar C = 0, \qquad s_d [b, \; \bar b, \; \phi, \; p_\lambda, \; \lambda] = 0, \nonumber\\
&& s_{ad} A_0 = -  C^\prime, \quad s_{ad} A_1 = - \dot C, \quad s_{ad} \Pi_\phi = - e \dot C, \qquad s_{ad} E = - \Box C, \nonumber\\ 
&& s_{ad} \bar C = -(\bar b - e \phi), \quad s_{ad} C = 0, \qquad s_{ad} [b, \; \bar b, \; \phi, \; p_\lambda, \; \lambda] = 0, \label{14}
\end{eqnarray}
which leaves (\ref{11}) invariant because of the fact 
\begin{eqnarray}
 s_d {\cal L}_d = - \partial_\mu [\bar b \; \partial^\mu {\bar C}], \qquad 
s_{ad} {\cal L}_d = - \partial_\mu [\bar b \; \partial^\mu  C].
\end{eqnarray}
Therefore, the corresponding action remains invariant under $s_{(a)d}$. 
At this stage, it is worthwhile to point out that the total gauge fixing term remains invariant under the (anti-)co-BRST symmetry transformations
(i.e. $s_{(a)d} [b (\dot A_0 - A_1^\prime) + \frac{1}{2} b^2] = 0$). This gauge fixing term has its origin in the co-exterior derivative 
$\delta = \pm * d *$ (with $\delta^2 = 0 $) of differential geometry as the operation of $\delta$ on a one-form produces the gauge-fixing term 
[i.e. $\delta A^{(1)} = (\partial \cdot A) $].
Here $*$ is the Hodge duality operation on the 2D spacetime manifold and the $\pm$ sign is dictated by the dimensionality of the 
spacetime \cite{egu,mukhi}. Thus, the nilpotent (anti-)co-BRST symmetry transformations has its origin to the co-exterior derivative ($\delta$) of 
differential geometry. 

Furthermore, it is straightforward to check that these (anti-)co-BRST symmetry transformations are absolutely anticommuting in nature
[i.e. $(s_d s_{ad} + s_{ad} s_d)\Phi = 0$] where $\Phi$ is any generic field of the theory.  \\

\noindent
{\bf 5. Bosonic symmetry transformations: Analogue of Laplacian operator}\\

\noindent
It is clear that the bosonized version of 2D VSM is endowed with four nilpotent (fermionic) symmetry transformations (i.e. $s_{(a)b}, s_{(a)d}$).
In addition to that, the following infinitesimal version of bosonic symmetry ($s_\omega = \{s_b, s_d \}$) transformations (with $s_\omega^2 \neq 0$)
\begin{eqnarray}
&& s_\omega A_0 = - b^\prime + \dot {\bar b} - e \dot \phi, \qquad s_\omega A_1 = - \dot b + {\bar b}^\prime - e \phi^\prime, \qquad s_\omega E = - \Box b, 
\nonumber\\
&&  s_\omega \Pi_\phi = - e(\dot b - {\bar b}^\prime + e \phi^\prime), \qquad  s_\omega [C, \; \bar C, \; \phi, \; p_\lambda, \; \lambda, \; b, \; \bar b] = 0, \label{16}
\end{eqnarray} 
also leaves the Lagrangian density (\ref{11}) quasi-invariant. It is explicitly given as follows
\begin{eqnarray}
 s_\omega {\cal L}_d =  \partial_\mu [e \; \varepsilon^{\mu\nu} \; \bar b \; \partial_\nu \phi -  e \; b \; \partial^\mu \phi]. 
\end{eqnarray}
The other anticommutator (i.e. $\{s_{ab}, s_{ad} \} = s_{\bar \omega}$) also produces a bosonic symmetry of the theory which is {\it not} 
independent of $s_\omega$. Moreover, it is easy to check that $(s_\omega + s_{\bar \omega}) \Phi = 0$, where $\Phi$ is any generic 
field of the theory. In summary, the following relationship is true
\begin{eqnarray}
 s_\omega \; = \; \{s_b, \; s_d \} \; = - \{s_{ab}, \; s_{ad} \} \; = \; - s_{\bar \omega}. 
\end{eqnarray}
The noteworthy point is that the ghost term  remains invariant under the bosonic symmetry transformations. This bosonic 
symmetry transformations find its analogue in terms of the Laplacian operator ($\Delta = \{d, \delta \}$) of differential geometry.  \\

\noindent
{\bf 6. Ghost and discrete symmetries}\\

\noindent
 The ghost number for the bosonic fields $A_0, A_1,$ $ \phi, b, \bar b, \lambda$ of the theory is equal to 
zero whereas the ghost number corresponding to the fermionic fields $C$ and $\bar C$ is equal to $\pm 1$. Thus, keeping above in mind, we define 
following ghost scale transformations: 
\begin{eqnarray}
&&  A_0 \to A_0, \qquad A_1 \to A_1, \qquad \phi \to \phi, \qquad b \to b, \qquad \bar b \to \bar b, \nonumber\\ 
&&  \lambda \to \lambda, \quad p_\lambda \to p_\lambda, \quad \Pi_\phi \to \Pi_\phi, \quad C \to e^{+ \Lambda} C, \quad \bar C \to e^{- \Lambda} \bar C.
\end{eqnarray}
In the above, $\Lambda$ is global infinitesimal scale 
parameter and $\pm 1$ in the exponentials of $C$ and $\bar C$ corresponds to the ghost numbers. The infinitesimal version
of the above mentioned ghost scale transformations $(s_g)$ can be given as
\begin{eqnarray}
&& s_g A_0 = 0, \quad s_g A_1 = 0, \quad s_g \phi = 0, \quad s_g b = 0, \quad s_g \bar b = 0,  \nonumber\\
&&  s_g \lambda = 0, \quad s_g p_\lambda = 0, \quad s_g \Pi_\phi = 0, \quad  s_g C = C, \quad s_g \bar C = - \bar C. \label{20}
\end{eqnarray}
These are the symmetry transformations as Lagrangian density (\ref{11}) remains invariant under $s_g$. 
Moreover, the ghost sector of Lagrangian density (\ref{11}) is also endowed with the following discrete symmetry transformations 
\begin{eqnarray}
 C \to \pm \; i \; \bar C, \qquad \bar C \to \pm \; i \; C. 
\end{eqnarray}
The above discrete symmetry transformations are useful in
enabling us to obtain the anti-BRST symmetry transformations from the BRST symmetries and vice versa. Furthermore, the above
transformation connects the co-BRST symmetry transformations to the anti-co-BRST symmetry transformations in the similar 
fashion as in the case of (anti-)BRST symmetry transformations. \\

\noindent
{\bf 7. Algebraic structures and physical relevance}\\

\noindent
 It is clearly shown, in previous sections, that the bosonized version of 2D VSM is endowed 
with the (anti-)BRST $(s_{(a)b})$, (anti-)co-BRST $(s_{(a)d})$, a bosonic symmetry $(s_\omega)$ and ghost scale symmetry $(s_g)$ 
transformations. The operator form of these symmetry transformations obey the following algebra
\begin{eqnarray}
&& s_{(a)b}^2 = 0, \quad s_{(a)d}^2 = 0, \quad \{s_b,\;  s_{ab} \} = 0, \quad  \{s_d,\;  s_{ad} \} = 0, \quad  s_\omega = (s_b + s_d)^2,  \nonumber\\
&& s_\omega  =  \{s_b, \; s_d \}  = - \{s_{ab}, \; s_{ad} \}, \qquad  [s_\omega, \; s_r] = 0, \quad r = b, ab, d, ad, g.  
\end{eqnarray}
These algebraic structure are exactly same as the algebra obeyed by the de Rham cohomological operators of differential geometry. 
The following algebra 
\begin{eqnarray}
& d^2 = 0, \qquad \delta^2 = 0, \qquad \Delta = \{d, \; \delta \} \equiv (d + \delta)^2, & \nonumber\\
& [ \Delta, \; d ] = 0, \qquad  [ \Delta, \; \delta] = 0,& 
\end{eqnarray}
is constituted by the de Rham cohomological operators, namely; exterior derivative ($d = dx^\mu \partial_\mu$), the co-exterior 
derivative ($\delta = \pm * d *$) and the Laplacian operator ($\Delta = (d + \delta)^2 = \{d, \delta \} $) of
differential geometry. Here $*$ is the Hodge duality operation on a manifold without boundary. 

Thus, on a compact manifold, we have following two-to-one mapping from the symmetries of the theory to the cohomological operators of 
differential geometry:  $(s_b, s_{ad}) \rightarrow d $, $(s_d, s_{ab}) \rightarrow \delta $ and  $ \{s_b, s_d \} = - \{s_{ab}, s_{ad} \} \rightarrow \Delta$. 
Hence, in this way, we can precisely identify all the symmetry transformations of the theory with the de Rham cohomological operators of differential geometry
where the latter are defined on a compact manifold without boundary (see, \cite{egu,mukhi,van} for details).  \\

\noindent
{\bf 8. Conclusions}\\

\noindent
The central theme of our present investigation was to obtain the off-shell nilpotent (anti-)co-BRST symmetry transformations
{\it together} with the usual (anti-)BRST symmetry transformations in the case of 2D bosonized version of vector Schwinger model. 
We have accomplished this goal. In fact, we have explicitly shown that the 2D bosonized version of VSM is endowed with, in totality, 
six continuous symmetry transformations as listed in (\ref{7}), (\ref{14}), (\ref{16}) and (\ref{20}). 

In our present model, the BRST symmetry transformations turn out to be the analogue of the exterior derivative of differential geometry 
as the kinetic term, having its origin to the exterior derivative, remains invariant under it. Similarly, the gauge fixing term, owing 
its origin to the co-exterior derivative, remains invariant under co-BRST symmetry transformations. Thus, co-exterior derivative 
can be realized in terms of co-BRST symmetries of the present theory. The anticommutator of BRST and co-BRST transformations 
produces a bosonic symmetry which is analogue of the Laplacian operator. 
It is the ghost terms of the theory which remain invariant under the bosonic symmetry transformations. 

Finally, we have shown that, at the algebraic level, the above mentioned symmetry transformations follow the same algebra as the 
algebra obeyed by the de Rham cohomological operators of differential geometry. It would be a nice endeavor to find the 
analogue of the Hodge duality operation $(*)$ in terms of {\it full} discrete symmetries of the present theory which, in turn, enable
us to prove this model to be a model for Hodge theory. This aspect 
is under investigation and our results will be reported in our future publications \cite{sgfu}.\\


\begin{thebibliography}{99}
\bibitem{brs1}  C. Becchi, A. Rouet and R. Stora, {\it Phys. Lett. B} {\bf 32}, 344 (1974)
\bibitem{brs2}  C. Becchi, A. Rouet and R. Stora, {\it Commun. Math. Phys.} {\bf 42}, 127 (1975)
\bibitem{brs3}  C. Becchi, A. Rouet and R. Stora, {\it Ann. Phys. (N. Y.)} {\bf 98},  287 (1976)
\bibitem{tyu}   I. V. Tyutin, Lebedev Institute Preprint, Report No: FIAN-39 \\ unpublished (1975)
\bibitem{sch1}  J. Schwinger,  {\it  Phys. Rev.} {\bf 128}, 2425 (1962)
\bibitem{suss}  A. Casher, J. Kougt and L. Susskind,  {\it Phys. Rev. Lett.} {\bf 31}, 792 (1973)
\bibitem{suss1} A. Casher, J. Kougt and L. Susskind,  {\it  Phys. Rev. D} {\bf 10},  732 (1974)
\bibitem{halp}  M. B. Halpern,  {\it  Phys. Rev. D} {\bf 13},  337 (1976)
\bibitem{boyan} D. Boyanovsky, I. Schmidt and M. F. L. Golterman, {\it  Ann. Phys.} {\bf 185},  111 (1988)
\bibitem{raja1} R. Jackiw and R. Rajaraman, {\it Phys. Rev. Lett.} {\bf 54}, 1219 (1985)
\bibitem{raja2} R. Rajaraman, {\it  Phys. Lett. B} {\bf 184}, 369 (1987)
\bibitem{flack} N. K. Falk and G. Kramer, {\it Ann. Phys.} {\bf 176}, 369 ( 1987)
\bibitem{malik} R. P. Malik, {\it Phys. Lett. B} {\bf 212}, 445 (1988)
\bibitem{usha}  U. Kulshreshtha, D. S. Kulshreshtha and H. J. W. M\"{u}ller-Kirsten, {\it Helv. Phys. Acta.} {\bf 66}, 752 (1993)
\bibitem{sght}  S. Gupta and R. P. Malik,  {\it Eur. Phys. J. C} {\bf 58}, 517 (2008)
\bibitem{sgcsm} S. Gupta, R. Kumar and R. P. Malik,  {\it Eur. Phys. J. C} {\bf 65}, 311 (2010)
\bibitem{malik1} S. Krishna, A. Shukla and R. P. Malik, {\it Mod. Phys. Lett.} A {\bf 26}, 2739  (2011) 
\bibitem{rkma3} R. Kumar, S. Krishna, A. Shukla and R. P. Malik, {\it Eur. Phys. J. C} {\bf 72}, 1980 (2012)
\bibitem{mal2p} R. Kumar, S. Krishna, A. Shukla and R. P. Malik, arXiv:1203.5519 [hep-th] 
\bibitem{green} M. B. Green, J. H. Schwarz and E. Witten, {\it Superstring Theory} Vols 1 and 2. (Cambridge: Cambridge University Press) (1987)
\bibitem{sgrr}  S. Gupta and R. P. Malik,  {\it Eur. Phys. J. C} {\bf 68},  325 (2010)
\bibitem{mal01} R. P. Malik, {\it J. Phys. A: Math. Gen.} {\bf 41},  4167 (2001)
\bibitem{mal03} R. P. Malik, {\it J. Phys. A: Math. Gen.} {\bf 36}, 5095 (2003)
\bibitem{mal04} R. P. Malik, {\it Mod. Phys. Lett.} A {\bf 14}, 1937 (1999)
\bibitem{yang}  H. S. Yang and B. H. Lee, {\it J. Math. Phys.} {\bf 37} 6106 (1996)
\bibitem{mcmu}  D. McMullan, {\it Commun. Math. Phys.} {\bf 149}, 161 (1992)
\bibitem{dirac} P. A. M. Dirac, {\it Lectures on Quantum Mechanics}. Belfer Graduate School of Science 
                (New York: Yeshiva University Press) (1964)
\bibitem{sunder} K. Sundermeyer, { \it Constrained Dynamics. Lecture Notes in Physics} (Berlin: Springer) Vol 169 (1982)
\bibitem{nem} D. Nemeschansky, C. Preitschopf and M. Weinstein, {\it Ann. Phys. (N.Y.)} {\bf 183}, 226 (1988)
\bibitem{egu} T. Eguchi, P. B. Gilkey and A. Hanson, {\it  Phys. Rep.} {\bf 66}, 213 (1980)
\bibitem{mukhi} S. Mukhi and N. Mukunda, {\it Introduction to Topology, Differential
Geometry and, Group Theory for Physicists} (New Delhi: Wiley Eastern) (1990)
\bibitem{van} J. W. van Holten, {\it Phys. Rev. Lett.} {\bf 64}, 2863 (1990)
\bibitem{sgfu} S. Gupta,  under preparation. 
\end{thebibliography}
\end{document}